\documentclass[nofootinbib,aps]{revtex4}
\usepackage{amsmath}
\usepackage{amssymb}
\usepackage[dvips]{graphicx}
\usepackage{psfrag}

\newcommand{\G}{\Gamma}

\newcommand{\Tr}{\mathrm{Tr}}

\newcommand{\cT}{{\mathcal T}}

\newcommand{\cO}{{\mathcal O}}

\newcommand{\cH}{{\mathcal H}}
\newcommand{\cM}{{\mathcal M}}

\newcommand{\cF}{{\mathcal F}}
\newcommand{\cK}{{\mathcal K}}
\newcommand{\cV}{{\mathcal V}}
\newcommand{\cZ}{{\mathcal Z}}
\newcommand{\cP}{{\mathcal P}}

\newcommand{\C}{\mathbb{C}}

\newcommand{\SO}{\mathrm{SO}}
\newcommand{\SU}{\mathrm{SU}}
\newcommand{\SL}{\mathrm{SL}}

\newcommand{\Ga}{\Gamma}
\def\bra{\langle}
\def\ket{\rangle}
\newcommand{\be}{\begin{equation}}
\newcommand{\ee}{\end{equation}}
\newcommand{\beq}{\begin{eqnarray}}
\newcommand{\eeq}{\end{eqnarray}}

\begin{document}
\title{\bf Group Field Theory: An overview\footnote{Prepared for the proceedings of
Peyresq Physics 9 Meeting: Micro and Macro Structure of Spacetime, Peyresq, France, 19-26 Jun 2004.}}
\author{L.Freidel\footnote{email:lfreidel@perimeterinstitute.ca}}
\affiliation{\hspace{1mm}
Perimeter Institute, 35 King Street North, Waterloo, Ontario, Canada N2J 2W9 \\
${}^b$\hspace{1mm} Laboratoire de Physique, \'Ecole Normale Sup\'erieure de Lyon,
46 all\'ee d'Italie, Lyon 69007, France}
\date{\today}

\begin{abstract}
We give a brief overview of the
properties of a higher dimensional generalization of matrix model
which arises naturally in the context of a background independent approach to
quantum gravity, the so called group field theory.
We show that this theory leads to a natural proposal for the physical scalar product of quantum gravity.
We also show in which sense this theory provides a third quantization point
of view on quantum gravity.
\end{abstract}
\maketitle

\section{Introduction}

\begin{flushright}{ \em ``Pluralitas non est ponenda sine neccesitate''}, William of Ockham (1285-1349).
\end{flushright}
Spin foam models describe the dynamics of loop quantum gravity in
terms of state sum models. The purpose of these models is to
construct the physical scalar product which is one of the main object of interest in quantum gravity.
Namely, given a $4$-manifold $M$ with boundaries $\Sigma_{0}$,
$\Sigma_{1}$ and given a diffeomorphism class of $3$ metric
$[g_0]$ on $\Sigma_0$ and $[g_1]$ on $\Sigma_1$ we want to compute
\be\label{phyprod}
\bra[g_0]|\cP|[g_1]\ket =
\int_{{\cal{M}}}{\cal{D}}[g] e^{i S(g)},
\ee
the integral being over $\cal{M}$: The space of all metrics on $M$ modulo 4-diffeomorphism
which agree with $g_0, g_1$ on $\partial M$.
The action is the Einstein Hilbert action and $\cP$ denotes the
projector on the kernel of the hamiltonian constraint.
This expression is of course highly formal,  there is no good non
perturbative\footnotemark \footnotetext{except in 2+1 dimension \cite{Witten3d}} definition of the
 measure on $\cal{M}$ and no good handle on the space of  kinematical states $|[g]\ket$.

In loop quantum gravity there is a good understanding of the kinematical Hilbert space
(see \cite{Ashtekar} for a review).
In this framework the states are given by {\it spin networks}
$\Ga_\jmath$ where
$\G$ is a graph embedded in a three space $\Sigma$ and $\jmath$ denotes a coloring of
the edges of $\Ga$ by representations of a group\footnotemark \footnotetext{In conventional loop quantum gravity the group is
$SL(2,\C)$, more generally $G$ is a Lorentz group} $G$ and a coloring of the vertex of $\Gamma$ by
intertwiners (invariant tensor) of $G$.
These states are eigenvectors of geometrical operators, the representations labelling
edges of the spin network are interpreted as giving a quanta of area $l_p^2
\sqrt{j(j+1)}$ to a surface intersecting $\Ga$.

In this context the spacetime is obtained as a spin network history:
If one evolve in time a spin network it will span a foam like structure i-e a combinatorial
2-complex denoted $\cF$.
The edges of the spin network will evolve into faces of $\cF$ the vertices of $\Ga$ will evolve into
edges of $\cF$ and transition between topologically different spin networks will occur at
vertices of $\cF$.
The spin network coloring induces a coloring of $\cF$: The faces of $\cF$ are colored by representation
$\jmath_f$ of $G$ and edges of $\cF$ are colored by intertwiners $\imath_e$ of $G$.
Such a colored two complex $\cF_{(\jmath_f,\imath_e)}$ is called a {\it spin foam} (\cite{Baez,Perez}).
By construction the boundary of a spin foam is an union of spin networks.

The definition is so far purely combinatorial, however if one restricts the 2-dimensional complex $\cF$ to be such that
$D$ faces meet at edges of $\cF$ and $D+1$ edges meet at vertices of $\cF$ we can reconstruct from
$\cF$ a $D$ dimensional piecewise-linear pseudo-manifold $M_\cF$ with boundary \cite{Depietri}.
Roughly speaking, each vertex of $\cF$ can be viewed to be dual to a $D$ dimensional simplex
and the structure of the 2-dimensional complex gives the prescription for gluing these
simplices together and constructing $M_\cF$.
The spin network states are dual to the boundary triangulation of $M_\cF$.

A local {\it spin foam model} is characterized by a choice
of local amplitudes $A_f(\jmath_f), A_e(\jmath_{f_e},\imath_e), A_v(\jmath_{f_v},\imath_{e_v})$ assigned respectively
 to the faces, edges and vertices of $\cF$.
$A_f$ depends only on the representation coloring the face, $A_e$ on the representations
of the faces meeting at $e$ and the intertwiner coloring the edge $e$, likewise
$A_v$ depends only on the representations and intertwiners of the faces and edges meeting at $v$.

Given a two complex $\cF$ with boundaries $\Ga_0,\Ga_1$ colored by $\jmath_0,\jmath_1$ the
Transition amplitude is given by
\be\label{spf}
A(\cF)=\bra\Ga_{\jmath_0}| \Ga_{\jmath_1}\ket_{\cF}
\equiv
\sum_{\jmath_f,\imath_e}
\prod_f A_f(\jmath_f)
\prod_e A_e(\jmath_{f_e},\imath_e)
\prod_v A_v(\jmath_{f_v},\imath_{e_v}),
\ee
the sum being over the labelling of internal faces and edges not meeting the boundary.
Note that a priori the amplitude depends explicitly on the choice of the two complex $\cF$.

There are many examples of such models.
Historically, the first example is due to Ponzano and Regge \cite{PR}:
They showed that the quantum  amplitude
for euclidean $2+1$ gravity with zero cosmological constant can be expressed as a spin foam model where
the group $G$ is $\SU(2)$, the faces are labelled by $\SU(2)$ spin $j_f$
\footnotemark \footnotetext{no edge intertwiner is needed since in 3d we restrict to only three face meeting along each edge
and there is a unique normalized intertwiner between three $\SU(2)$ representation.}
and the local amplitudes are given by
$A_f(\jmath_f)= (2j_f+1)$, $A_e(\jmath_{f_e})=1$
and the vertex amplitude $A_v(\jmath_{f_v})$ which depends on $6$ spins is the normalized $6j$ symbol
or Racah-Wigner coefficient.
The remarkable feature of this model is that it doesn't depend on the choice of the two complex $\cF$
but only on $\cM_{\cF}$.
The inclusion of a cosmological constant or the description of lorentzian gravity can be implemented
easily by taking the group
to be a quantum group \cite{TV} or to be a non compact Lorentz group \cite{Lor}.
Along the same line, it was shown  that 4d topological field theory
called $BF$ theory \cite{BF} can be quantized in terms of
triangulation independent spin foam model \cite{Ooguri}.

It was first realized by M. Reisenberger that spin foam models give
a natural arena to deal with 4d quantum gravity
\cite{Reisenberger}.
Two seminal works triggered more interest on spin foam models.
In the first one, Barrett-Crane \cite{BarrettCrane} proposed a spin
foam model for 4d general relativity\footnotemark \footnotetext{more precisely a
prescription for the vertex amplitude.}. This model is obtained
from the spin foam model of pure $BF$ by restricting the
Lorentz representations to be simple\footnotemark \footnotetext{If we label the
representation of $\SL(2,\C)$, par a pair of SU(2) spins $(j,k)$, the simple representations
are the ones where $j=k$.} so that the spin coloring the
faces are $SU(2)$ representations in the Euclidean context.
 In the second one, it was shown by Reisenberger and Rovelli
\cite{RR} that the evolution operator in loop quantum gravity can
be expressed as a spin foam model and they propose an
interpretation of the vertex amplitude in terms of the matrix
elements of the hamiltonian constraint of loop quantum gravity
\cite{Th}. The spin labelling the faces are also $SU(2)$ representations
and are interpreted as quanta of area.
this construction has been exemplified in 2+1 gravity \cite{PN}

It was soon realized that spin foam models can naturally
incorporate causality \cite{CausalFL}, Lorentzian  signature \cite{Lor2}
and coupling to gauge field theory \cite{gauge}.
The Barrett-Crane prescription was understood to be linked to the
Plebanski formulation of gravity where the Einstein action is
written as a $BF$ theory subject to constraints \cite{DePietriL}.
This formulation and the corresponding spin foam models were
extended to gravity in any dimensions \cite{PuzioKL}.

The main lesson is that spin foams are  gives a very general framework
which allows to address in a background independent manner the
dynamical issues of a large class of diffeomorphism invariant
models including gravity in any dimensions coupled to gauge fields
\cite{KLClassical}. This formulation naturally incorporated the
fact that the kinematical Hilbert space of the theory is labelled
by spin networks\footnotemark \footnotetext{This is relevant in view of the `LOST'
uniqueness theorem stating that there is a unique diffeomorphism invariant
representation \cite{LOST} of a theory with phase space a pair of
electric and magnetic field.}.

A different line of development originated from the detail study
of the vertex amplitude proposed by Barrett-Crane and the
corresponding higher dimensional quantum gravity models
\cite{Barrett, FeynmanKL}. These studies shows that these amplitudes
can be written  as some Feynman graph evaluation.
For instance in the original Barrett-Crane model
\be\label{BC}
A_v(\jmath_1,\cdots,\jmath_{10})= \int_{S^3} \prod_{i=1}^{5}dx_i
\prod_{i\neq j}G_{\jmath_{ij}}(x_i,x_j),
\ee
where the ten spins are simple representations of $SO(4)$ labelling the $10$ faces of the 4-simplex and
$G_\jmath(x,y)$ is the Hadamard propagator of $S^3$, $(\Delta_{S^3}+
j(j+1))G_j =0$, $G(x,x)=1$.

This structure was calling a field theory interpretation of spin foam models.
It was eventually found in
\cite{DPRKL} that  the Barrett-Crane spin foam model can remarkably be interpreted as a Feynman graph of a new
type of theory baptized `group field theory' (GFT for short).
The GFT structure was first discovered by Boulatov \cite{Boulatov} in the context of three dimensional gravity
where a similar connection was made and further developed by Ooguri
in the context of $4$d $BF$ theory \cite{Ooguri}.
Ambjorn, Durhuus and Jonnson \cite{ADJ} also pointed out similar structure in the context of dynamical triangulation.
It is clear in this context that group field theory can be understood in a precise sense as a higher
dimensional generalization of matrix models which generate a summation over $2d$ gravity models \cite{2drev}.

Reisenberger and Rovelli \cite{RR2}  showed, in a key work, that the appearance of GFT in
the context of spin foam models is not an accident but a generic feature.
They proved that {\it any} local spin foam model of the form (\ref{spf}) can be
interpreted as a Feynman graph of a group field theory.

We have argued that spin foam models generically appear in the context of
background independent approach to quantum gravity \cite{KLClassical}, this result shows that
GFT is an important and unexpected universal structure behind the dynamics of such models.
A deeper understanding of this theory is clearly needed.
GFT was originally designed to address one of the main shortcomings of the spin foam approach: namely the fact that
the spin foam amplitude (\ref{spf}) depends explicitly on the discrete
structure (the two complex or triangulation).
As we will now see in more details it does much more than that and give a
third quantization point of view on gravity where spacetime is emergent
and dynamical.

\section{Group field theory}
\subsection{definition}

In this section we introduce the general GFT action that can be
specialized to define the various spin foam models described in
the introduction.

We consider a Lie group $G$ which is the Lorentz group in
dimension $D$ ($G=SO(D)$ for Euclidean gravity models and
$G=SO(D-1,1)$ for Lorentzian ones \footnotemark \footnotetext{It
is also possible to generalize the definition to quantum groups or
fuzzy group, we will restrict the discussion to compact group to
avoid unnecessary technical subtleties.}. $D$ is the dimension of
the spacetime and we will call the corresponding GFT a $D$-GFT.
 The {\it field}
$\phi(x_1,\cdot,x_D)$, denoted $\phi(x_i)$ where $i=1\dots D$,
is a function on $G^D$.
The dynamics is defined by an action of the general form
\begin{eqnarray}\nonumber
    S_D[\phi] = \frac{1}{2} \int
dx_i \, dy_i \ \phi(x_i)\, \cK(x_iy_i^{-1})\, \phi(y_{i})
  + \frac{\lambda}{D+1} \int \prod_{i\neq j =1}^{D+1} dx_{ij}  \ \cV(x_{ij}x_{ji}^{-1})
~\phi(x_{1j})\cdots \phi(x_{D+1j}) ,
\label{actionexpanded}
\end{eqnarray}
where  $dx$ is an invariant measure on $G$, we use the notation
$\phi(x_{1j}) =\phi(x_{12},\cdots,x_{1D+1})$.
 $\cK(X_i)$ is the kinetic   and  $\cV(X_{ij})$ ($X_{ij}=x_{ij}x_{ji}^{-1}$) the interaction
 kernel, $\lambda$ a coupling constant, the interaction is chosen to be homogeneous of degree $D+1$.
 $\cK,\cV$ satisfy the
 invariance properties
 \be\label{inv}
 \cK(gX_{i}g^\prime) = \cK(X_{i}),\quad \cV(g_{i}¥X_{ij}g_{j}^{-1})= \cV(X_{ij})\quad \forall g,g',g_i \in G.
 \ee
 This implies that the action is invariant under the gauge
 transformations
 $\delta\phi(x_{i}) = \psi(x_{i})$,
 where $\psi$ is {\it any} function satisfying
 \be \label{sym1}
 \int_{G} dg \psi(gx_{1},\cdots,gx_{D}) =0.
 \ee
 This symmetry is gauge fixed if one restricts the field $\phi$ to
 satisfy $\phi(gx_{i})=\phi(x_{i})$.
The action is also invariant under the global symmetry
\be\label{sym2}
\phi(x_{1},\cdots,x_{D})\rightarrow \phi(x_{1}g,\cdots,x_{D}g).
\ee
 The main interest of these theories resides in the following crucial
 properties they satisfy. Most of them are well established, some
 are new (property 4) and some (property 6) still conjectural.
 Altogether they give a picture of the relevance of GFT for background
 independent approach to quantum gravity and lead to the conclusion (more precisely the conjecture) that
 GFT provides a third quantization of gravity.
\vspace{4mm}

{\bf GFT properties:}~\begin{enumerate}
 \item\label{1} The Feynman graphs of a D-GFT are cellular complexes $\cF$ dual to a $D$ dimensional
triangulated topological spacetime $M_\cF$.
\item\label{2}
The Feynman graph evaluation of a GFT is equal to the spin foam amplitudes of a local spin foam model.
Conversely, any local spin foam model can be obtained from the evaluation of a  GFT Feynman graph.
\item Spin networks label polynomial gauge invariant operators of the GFT.
\item The tree level two-point function of GFT gauge invariant operators
gives a proposal for the physical scalar product. This proposal
involves spacetime of trivial topology and is triangulation independent.
\item The full two-point function of GFT  gauge invariant operators gives
a prescription for the quantum gravity amplitude including a sum over all topologies.
\item The possible loop divergences of GFT Feynman graphs are interpreted to be a consequence of a residual
action of spacetime diffeomorphisms on spin foam.
One expect a relation between the renormalization group of GFT and
the group of spacetime diffeomorphisms.
\end{enumerate}

\subsection{GFT: Examples and Properties}
In this section we give some examples and illustrate the properties listed
above.

{\bf Some examples}
The simplest examples comes from the choice
\be\label{ex}
\cK(x_{i},y_{i}¥) =\int_G dg \prod_{i} \delta(x_{i}y_{i}^{-1}g),\quad
\cV(X_{ij}) =\int \prod_i dg_i
\prod_{i<j} \delta(g_iX_{ij}g_j^{-1}),
\ee
$\delta(\cdot)$ is
the delta function on $G$ and the integrals insure the gauge invariance (\ref{sym1}).

If one further restricts to dimension $D=2$ the
symmetry property implies that $\phi(x_1,x_2)=\tilde{\phi}(x_1^{-1}x_2)$.
$\tilde\phi$, being a function on the group, can be expanded in Fourier modes.
Lets consider $G=SU(2)$, denote by $V_j$ the spin $j$ representation,
$d_j$ its dimension, $D^j(x)\in End(V_j) $ the group matrix element and define
\be\label{fourier}
\Phi_j \equiv \int dx\,  \tilde{\phi}(x) D^j(x^{-1}),\quad
 \tilde{\phi}(x)=\sum_j d_j \Tr\left(\Phi_j D^j(x)\right).
 \ee
One can readily see \cite{CL} that the GFT reduces to a sum of matrix models:
\be
 S_2[\phi] = \sum_j d_j \left(\Tr(\Phi^2_j) + \frac{\lambda}{3}\Tr(\Phi_j^3)\right).
\ee
It is well known that the Feynman graph expansion of a matrix model
is expressed in terms of fat graphs \cite{2drev}, each edge can be represented as a double line
each one carrying a matrix index, the trivalent interaction implies that this graph is dual to a
triangulation of a two dimensional closed surface.
Moreover if one computes the Feynman evaluation of a genus $g$ diagram $\Gamma$ diagram we find
\be
I(\Gamma) = \sum_j d_j^{2-2g},
\ee
which is the evaluation of the partition function of topological $BF$ theory in 2 dimension
on a surface of genus $g$\footnotemark. \footnotetext{If one choose the propagator
$\cK(x_{i}¥,y_{i}¥) =\int_G dg \prod_{i} \delta_\epsilon(x_{i}y_{i}^{-1}g)$ with
$\delta_\epsilon$ the heat kernel on the group, $(\partial_\epsilon +\Delta_G)\delta_\epsilon =0, \delta_0(g)
=\delta(g)$ we obtain the partition function of 2D Yang-mills theory as a Feynman graph evaluation \cite{Witten}}
This result exemplifies the properties (\ref{1},\ref{2}).
This property generalizes to any $D$, the Feynman graph evaluation of the example (\ref{ex})
gives the partition function of $BF$ theory in dimension $D$.

\begin{figure}
\begin{center}
\includegraphics[height=3cm]{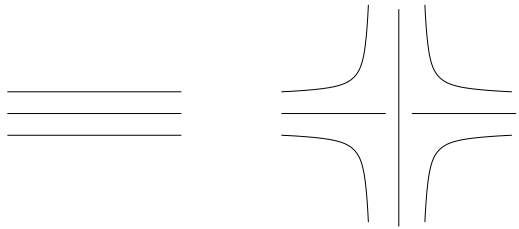}
\end{center}
\caption{Graphical representation of the propagator and
interaction of a $3$-GFT}
\label{Frules}
\end{figure}

{\bf Property 1.} To illustrate this property in higher dimension lets consider the case of dimension $D=3$, the field
$\phi$ possesses three arguments, so each edge of a Feynman graph possesses three strands running parallel to it,
$4$ edges meet at each vertex and the form of the interaction $\cV$ forces the strands to recombine as in
figure (\ref{Frules}).
Each strand of the graph forms a closed loop which can be interpreted as the boundary of a $2d$ disk.
These data are enough to reconstruct a topological 2d complex $\cF$, the vertices and edges of this complex
correspond to vertices and edges or the Feynman graph, the boundary of the faces of $\cF$ correspond to
the strands of the Feynman graph.
As we have already emphasized  we can reconstruct a triangulated $3d$ pseudo-manifold
from such data.
\begin{figure}[t]
\begin{center}
\includegraphics[height=3cm]{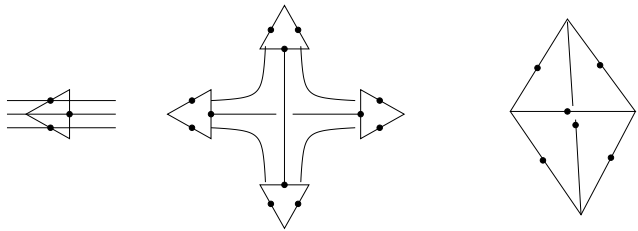}
\end{center}
\caption{Triangulation generated by Feynman diagrams}
\label{tet}
\end{figure}
It can be understood as follows: The three strands running along
the edges can be understood to be dual to a triangle and the
propagator gives a prescription for the gluing of two triangles.
At the vertex $4$ triangles meet and their gluing form a
tetrahedra (see figure (\ref{tet}). With this interpretation the
Feynman graph of a GFT is clearly dual to a three dimensional
triangulation. This is true in any dimension \cite{Depietri,DPRKL}.

This means that the perturbative expansion of the partition function
can be expressed as a sum over two complexes
\be\label{exp}
\cZ =\int {\cal D}\phi e^{-S_D[\Phi]} = \sum_\cF \frac{\lambda^{|V|}}{\mathrm{sym(\cF)}} I(\cF),
\ee
where the sum is over two complexes, $|V|$ is the number of vertices of $\cF$, ${\mathrm{sym(\cF)}}$ the
symmetry factor of $\cF$ and $I(\cF)$ the GFT Feynman graph evaluation of the complex $\cF$.

{\bf Property 2.} Since the field $\phi$ is a function on $D$ copies of the group it can be expanded in Fourier
modes (\ref{fourier}), the `momentum' of this field are  spins which circulate along the
strands of the Feynman graph or equivalently which label the faces of the complex $\cF$.
From the quantum gravity point of view this spin is interpreted as a quanta of area carried by the face.
The computation of the Feynman graph in `momentum' space contains contributions from the
Fourier modes of the propagator which gives edges amplitude $A_e$; from the
Fourier mode of the interaction kernel which gives a vertex amplitude $A_v$ and from the trace over
the representation circulating on each face. Overall, the evaluation $I(\cF)$ can be expressed as a
local spin foam model (\ref{spf}).

In order to make this correspondence more precise lets
give a geometrical interpretation of the invariance property (\ref{inv}) of the
interaction kernel $\cV$: Let $\Gamma_{D}$ be the graph of  a $D$ dimensional simplex
which consists of $D+1$ vertices and $D(D+1)/2$ edges.
$\cV$ is a function of group elements associated with the edges
of $\Gamma_{D}$ which is invariant under an action of the gauge group
at the vertices of $\Gamma_{D}$. We denote the space of such function
by $L^{2}(\Gamma_{D})$.  $L^{2}(\Gamma_{D})$ admits an orthonormal
basis labelled  by spin networks $(\Gamma_{D},\jmath_{ij},\imath_{i})$, where
$\jmath_{ij}$ are spins labelling the edges  and $\imath_{i}$ are intertwiners
labelling the vertices of $\Gamma_{D}$. Given $(\Gamma_{D},\jmath_{ij},\imath_{i})$,
we can uniquely construct a spin  network functional
$\Theta_{\jmath_{ij},\imath_{i}}(X_{ij})$\footnotemark
\footnotetext{\label{spnetdef}Given a spin network $(\Ga,\jmath_{e},\imath_{v})$
the spin network functional $\Theta_{\Ga,\jmath_{e},\imath_{v}}(x_{e})$
is obtained by contracting the matrix elements of $x_{e}$ in the
representation $\jmath_{e}$ with the intertwiners $\imath_{v}$
according to the topology of the graph $\Ga$. By construction this
functional is invariant under the action of the group at each vertex
of $\Ga$.}.
The interaction kernel can be expanded in terms of this basis as follows
\be\label{vertex}
A_v(\jmath_{ij},\imath_{i})=
\int \prod_{i<j}dX_{ij}\cV(X_{ij})
\Theta_{{\jmath_{ij},\imath_{i}}}(X_{ij}),\quad
\cV(X_{ij}) = \sum_{\jmath_{ij},\imath_{i}} \prod_{i<j}d_{\jmath_{ij}} A_v(\jmath_{ij},\imath_{i})
 \Theta_{{\jmath_{ij},\imath_{i}}}(X_{ij}).
 \ee
Similarly, the quadratic kernel $\cK(X_{i})$ can be expanded in terms of
the spin network functional $\Phi_{\jmath_{i}}(X_{i})$ associated
with the `theta' graph which consists of two vertex joined by $D$
edges:
\be\label{edge}
1/A_e(\jmath_{i})=
\int \prod_{i}dX_{i}\cK(X_{i})
\Theta_{\jmath_{i}}(X_{i}) \quad \cK(X_{ij}) = \sum_{\jmath_{i}} \prod_{i}d_{\jmath_{i}}
1/A_e(\jmath_{i})
 \Theta_{{\jmath_{i}}}(X_{i}).
\ee
It is now a direct computation to show that $I(\cF)$ is expressed as a
local spin foam model with the edges and vertex amplitude determined
by (\ref{vertex},\ref{edge}) and the face amplitude being the
dimension of the representation labelling the face\footnotemark
\footnotetext{It is possible to produce different face amplitude by modifying the
symmetry properties of the action \cite{DPRKL}.}. Conversely, given a
spin foam model we can reconstruct a GFT via (\ref{vertex},\ref{edge}).
This establishes the equivalence or duality between spin foam models and GFT, which was first
proven in \cite{RR2}.
One can now check that the example (\ref{ex}) gives in dimension $3$
the Ponzano-Regge model and in higher dimensions the discretization of
topological $BF$.
With this prescription we can also reconstruct from the Barrett-Crane amplitude (\ref{BC})
the interaction kernel:
\be\label{BCker}
\cV_{BC}(X_{ij}) = \int_{G}\prod_{i}dg_{i} dh_{i} \int_{H}
\prod_{i\neq j} du_{ij} \prod_{i<j}
\delta(g_iu_{ij}h_{i}X_{ij}(g_ju_{ji}h_{j})^{-1}),
\ee
where $G=\SO(4)$ and $H=\SO(3)\subset \SO(4)$.

{\bf Property {3}.} We have discussed so far only the partition
function of the GFT, but one should also consider expectation
values of GFT operators. The physical operators $\cO(\phi)$ should
be gauge invariant under (\ref{sym1},\ref{sym2}). Such operators
can be constructed with the help of spin network: Lets consider a
spin network $(\Ga,\jmath_{e},\imath_{v})$ such that all its
vertices have valency $D$. and lets denote
$\Theta_{(\Ga,\jmath_{e},\imath_{v})}(x_{e})$ the corresponding
spin network functional (see footnote \ref{spnetdef}). We denote by
$V_{\Ga}$, $E_{\Ga}$ the set of vertices of $\Ga$ and define the
observable of the $D$-GFT \be
\cO_{(\Ga,\jmath_{e},\imath_{v})}(\phi) = \int_{G} \prod_{(ij)\in
E_{\Ga}} dx_{ij} dx_{ji}
\Theta_{(\Ga,\jmath_{e},\imath_{v})}(x_{ij}(x_{ji})^{-1})
\prod_{i\in V_\Ga} \phi(x_{ij}). \ee The element $x_{ij}$ are
associated to all the edges of $\Ga$ meeting at the vertex $i$, by
construction there is always $D$ such elements. This observable
is homogenous in $\phi$, the degree of homogeneity being the
number of vertices of $\Ga$. It is straightforward to check that
this observable respect the symmetries of the GFT.

{\bf Property 4.}
We can now come back to the original problem, that is the construction
of the physical scalar product. Since spin networks label operators of
the GFT, we propose to define this scalar product as the evaluation of
the GFT two point function in the {\it tree level} truncation.
Namely, given two $D$ valent spin network $\Ga_{1}$, $\Ga_{2}$
having $N_{1}, N_{2}$ vertices we define
\be \label{phyprod2}
\bra \Ga_{1}|\Ga_{2}\ket_0 \equiv \bra \cO_{\Ga_{1}}|\cO_{\Ga_{2}}\ket_{Tree}
= \sum_{\cF \in \cT_{N_{1},N_{2}}}  \frac{I(\cF)}{\mathrm{sym(\cF)}}.
\ee
$\cT_{N_{1},N_{2}}$ denote the space of GFT Feynman graphs supported on
 connected trees having $N_{1}$ initial univalent vertices and $N_{2}$ final univalent vertices, we
sum over all of them.

This simple proposal for the scalar product does not depend on a
particular triangulation, and therefore it addresses one of the main shortcomings of the spin
foam approach. A related idea has been already been considered by Gambini and Pullin in a different context \cite{GP}.
This product satisfies two crucial properties: First, it is well defined
and finite, since it is a tree level evaluation no infinite summation
are involved. Second, it is positive but
not strictly positive, it possesses a kernel.
This kernel should be expected, since the physical scalar product of
quantum gravity (\ref{phyprod}) computes the matrix elements of the
projector on the kernel of the hamiltonian constraints.
This means that any vector in the image of the hamiltonian constraint
belongs to the kernel of (\ref{phyprod}).
In our case, one can show that the GFT scalar product
(\ref{phyprod}) has a kernel which is in the image of the GFT
equation of motion.
Namely, the following gauge invariant observable
\be \label{kerob}
\delta \cO_{(\Ga,\jmath_{e},\imath_{v})}(\phi) =
\int_{G} \prod_{i}dx_{i}
 \left(\cK^{-1} \frac{\delta S[\phi]}{\delta \phi} \right)(x_{i})
 \frac{\delta \cO_{(\Ga,\jmath_{e},\imath_{v})}(\phi)}{\delta
 \phi(x_{i})},
 \ee
 which is proportional to the GFT equation of motion, is in the kernel of (\ref{phyprod2}).
 In this formula $\cK^{-1}$ is the propagator, it is convoluted with the
 equation of motion.
 We can expand this observable as a linear combination of spin
 network observables, the first term in the expansion of
 $\delta \cO_{\Ga}$ is $\cO_{\Ga}$ the other terms are spin network
 observables containing $D$ more fields.

 The physical Hilbert space can be constructed as an application of the Gelfand-Naimark-Segel theorem.
 It is obtained  from the kinematical Hilbert space
 spanned by spin networks,  by quotienting out
 the vectors in the kernel of (\ref{phyprod2}):
  $\cH_{\mathrm{phys}} = \cH/{\mathrm{Ker}\bra \cdot|\cdot\ket_0}$ \cite{RP}.
  The induced scalar product on $\cH_{\mathrm{phys}}$ is positive definite.

 The product (\ref{phyprod2}) involves only tree Feynman graphs. Using
 the correspondence between GFT Feynman graphs and discrete manifolds
 one sees that all the manifolds involved in the sum are of the same
 topology and describe a ball on the boundary of which the operators
 are inserted.
 Since this product (\ref{phyprod2}) is independent of the choice of
 triangulation it can be thought as a `continuous' scalar product.
 One might worry that this is realized without taking any sort of
 continuum or refinement limit and therefore that this prescription
 describes some sort of topological field theory.
 This is not the case, in this prescription the complexity of the
 spacetime triangulation involved in the summation grows with the
 complexity of the boundary spin network state. If one considers
 highly complicated spin network states that approach a continuum geometry,
 the corresponding spin foams (discrete spacetimes) involved in the summation
 are also highly complicated and good approximation of a continuous
 geometry.
 Also, we have seen that $\delta \cO_{\Ga}$  is a linear combination
 of $\cO_{\Ga}$ and higher order spin network operators.
 If we think of $\Gamma$ as being dual to a space triangulation, the higher order
 spin network operators are dual to a refined triangulation. We can
 therefore replace in the computation of the scalar product, the state
 ${\Ga}$ by a linear combination of spin network states of higher
 degree, and continue this replacement in each new terms ad infinitum, therefore
 ending with a expression for the scalar product in terms of
 a linear combination of arbitrarily fine triangulations which gives a `true'
 continuum limit expression of the scalar product.
 This use of the kernel expresses the fact that a subset of the Hilbert space based on
 a fine triangulation can be effectively described in terms of states
 leaving on a coarser one. The theory is not topological if this
 subset is a proper subspace (this is the case for the Barrett-Crane
 model for instance).

This proposal for a physical scalar product in the context of loop
quantum gravity closes in some sense a long quest starting from the
construction of the hamiltonian constraints, the search for its solutions
and  the construction of the physical scalar product. It gives us  a hint on
the answer to the last question (which contains the others), it doesn't
end the quest but provides a new starting point. The problem is now to
understand the dynamical content which is contained in such a proposal
and to see whether at least one GFT (the Barrett-Crane one for instance)
possesses the right dynamical content and can reproduce the physics of
general relativity in the infrared. This question is not new, but with
the help of the GFT it can be asked for the first time in terms of a
proposed physical scalar product.
The difficulty resides in the fact that the dynamics is encoded in
terms of spin networks transition amplitudes, a language far remote from semiclassical physics and one
needs to design criteria to select the right model or to test and
eventually refute a proposed one.

\begin{figure}
\psfrag{+}{$+\,\, \sum_jA_v(j)$} \psfrag{j}{$j$}\psfrag{=}{$=$}\psfrag{a}{$\alpha^2$}
\includegraphics{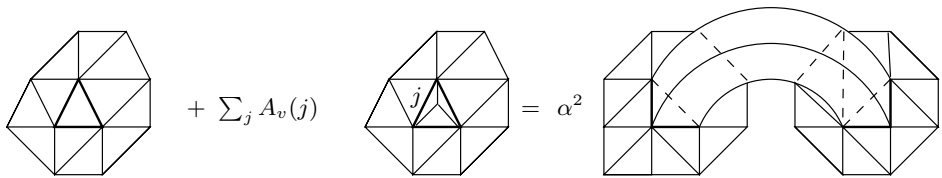}
\caption{A graphical representation of the Schwinger-Dyson equation in terms of triangulation.}
 \label{schwinger}
\end{figure}

{\bf Property 5.}
The previous scalar product can be naturally extended to include a sum over all Feynman graphs
of the GFT, this was the original proposal \cite{DPRKL}, the gravity amplitude is in this case
\be\label{fullexp}
\bra \Ga_{1}|\Ga_{2}\ket_\lambda=
\ln\left[\frac{\lambda^{-\frac{N_1+N_2}{D-1}}}{\cZ}\int {\cal D}\phi\,
\cO_{\Ga_{1}}(\phi)\cO_{\Ga_{2}}(\phi)e^{-S_D[\Phi]}\right] =
\sum_{\cF, \partial\cF =  \Ga_{1}\cup \Ga_{2}}
 \frac{\lambda^{|V|-\frac{N_1+N_2}{D-1}}}{\mathrm{sym(\cF)}} I(\cF).
\ee
where $|V|$ is the number of vertices of $\cF$, this correspond to the number of
$D$ simplex of the dual triangulation, the sum is over connected graph matching the
given spin network on the boundary, $N_1, N_2$ are the degree of homogeneity of the operators
$\cO_{\Ga_1},\cO_{\Ga_2}$ and $\cZ$ is the partition function (\ref{exp}).
The coupling parameter $\lambda$ weights, in the perturbation expansion, the
size of the discrete spacetime. Indeed its power in the perturbative expansion is the
 number of D-dimensional simplices, which can be understood as a `spacetime volume' in discrete units.
  It can be given another interpretation:
Lets define $\alpha= \lambda^{1/D-1}$ and lets redefine the fields
$\tilde{\phi} = \alpha \phi$, the action becomes
$S_\lambda[\phi]= {1}/\alpha^2 \tilde{S}[\tilde{\phi}]$ where $\tilde{S}=S_{\lambda=1}$ is independent of
the coupling constant.
The amplitude can be expanded in $\alpha$,
$\bra \Ga_{1}|\Ga_{2}\ket_\alpha=\alpha^{-2}\sum_i \alpha^{2i} \bra \Ga_{1}|\Ga_{2}\ket_i$
where $ \bra \Ga_{1}|\Ga_{2}\ket_i$ is a sum of GFT Feynman graphs containing $i$ loops.
From the space time point of view adding a loop to a GFT Feynman diagram amounts to adding a handle
to the discrete manifold. Hence $\alpha$ controls the strength of topology change. In the limit $\alpha=0$,
we recover the classical evaluation (\ref{phyprod2}) where topology change is suppressed.
This can be also understood by looking at the Schwinger-Dyson equation of motion.
Lets focus for simplicity on the nucleation amplitude where $\Ga_1$ is empty so that $\cO_{\Ga_1}=1$
which describes the creation of a spacetime from nothing.
The Schwinger-Dyson equation reads (see figure \ref{schwinger})
\be\label{SD}
\bra \delta \cO_{\Ga}\ket_\alpha = \alpha^2 \bra \delta^2 \cO_{\Ga}\ket_\alpha,
\ee
where $\delta \cO_{\Ga}$ (\ref{kerob}) is in the kernel of the physical scalar product
$\bra\cdot|\cdot\ket_0$ as can be easily seen from taking the limit $\alpha  \rightarrow 0$ in
(\ref{SD}).
and
\be
\delta^2 \cO_{\Ga} \equiv \int_G \prod_{i}dx_i dy_j \cK^{-1}(x_i,y_j)
\frac{\delta^2 \cO_{\Ga}(\phi)}{\delta
 \phi(x_{i})\delta \phi(y_{j})}.
 \ee

 We have seen that the field $\phi$ is dual to a $D-1$ simplex, the operator $\delta\cO_\Ga$
 corresponds to a sum of spin network boundary states, one being a triangulation dual to $\Ga$ the others
 obtained by subdividing one of the $D-1$ simplex of this triangulation.
 The operator $\delta^2 \cO_{\Ga}$ deletes two $D-1$ simplices and glues the resulting
 holes together thus creating a handle. This handle creation is weighted by $\alpha^2$.

 Note that the scalar product $\bra\cdot|\cdot\ket_\alpha$ is  strictly positive and
  doesn't possess any non trivial kernel since
 $\bra\cO|\cO\ket_\alpha$ is the integral of a positive quantity\footnotemark \footnotetext{Of course this
 integral is well defined only if we introduce a cutoff on the `momentum' (spin) of the field. So more
 work is needed in each particular model to prove that such a statement is valid in the unregularised theory.}.
The states $\delta \cO_\Gamma$ which are in the kernel of $\bra \cdot| \cdot \ket_0$
now generates topology change (\ref{SD}).

Since we now include Feynman graphs with loops we have to worry about potential perturbative
divergences due to the Feynman graph evaluation.
A careful analysis shows that the potential divergences of the GFT are not associated with loops but with
higher dimensional analogs: The so called `bubbles' of the spin foam \cite{perez1}.
A bubble is a collection of faces of the 2 complex $\cF$ which forms a closed surface. Each time a bubble appears
the sum over spins (GFT momenta) is unrestricted and potentially infinite.
A remarkable finiteness result was proven in \cite{perez2} for the Barrett-Crane model.
It was shown that if one takes the interaction kernel (\ref{BCker}) and the propagator (\ref{ex}),
there are no divergences arising in the computation of Feynman graph (associated with regular 2d complex),
the corresponding GFT is super-renormalizable in this case.

There is also a possibility of potential non-perturbative divergences which arises from the sum over topology.
It is well known that the number of triangulated manifold of arbitrary topology grows factorially with the
number of building block and the sum (\ref{fullexp}) is therefore not convergent for any non zero value of
$\alpha$. This is not surprising from the GFT point of view since we know that a perturbative expansion should be
interpreted as an asymptotic series, not a convergent series.
In some cases, this series is uniquely Borel summable and the GFT provides a non perturbative definition for the sum over
all topologies. This was shown to be true in the context of $3d$ gravity where Borel summability of a
mild modification of the Boulatov model was proven \cite{FLnonpert}.
It is not known whether this can be achieved in $4d$ gravity.

{\bf Property 6}
As we already mentioned, the key open problem is to gain an understanding of the low energy effective
physics from the GFT. One proposal for addressing this question is to focus on the issue of the diffeomorphism
symmetry. We know that any theory which depends on a metric reproduces usual gravity plus higher derivative
 corrections in the infrared if it is invariant under spacetime diffeomorphism.
In loop quantum gravity,  spin networks label the gravitational degree of freedom, this suggests that
any spin foam model which can be shown to respect spacetime diffeomorphism will contain
gravity in a low energy limit.
The problem is therefore to have a proper understanding of the action of spacetime
diffeomorphism on spin foam models.
It was argued in \cite{Ldiff} (and exemplified in the context of 3d gravity) that
diffeomorphism symmetry should act as a gauge symmetry on the spin foam amplitudes.
This means that the initial spin foam amplitudes \ref{spf} which are not gauge fixed with respect
to this symmetry and which do not break diffeomorphism symmetry should possess divergences
coming from the ungauged integration over the diffeomorphism gauge group.
Diffeomorphism symmetry is due to the Bianchi identity which is a three form on space time
and then couples to the bubbles of spin foams.
Therefore diffeomorphism symmetry should manifest itself in the bubble divergences.

This analysis leads to the conclusion that the Barret-Crane GFT model proposed in \cite{perez1,Oriti}
which has no bubble divergences is not a satisfactory model\footnotemark. \footnotetext{The prescription
of \cite{perez1,Oriti} differs from the original prescription \cite{DPRKL} (which possess bubble divergences) by the choice of the kinetic term
of the GFT. This kinetic term controls the way different vertex amplitude are glued together.
There is a large consensus and good understanding of the Barrett-Crane vertex amplitude but so far,
no general agreement on the choice of the kinetic term has been reached.
Different choices leads to different properties with respect to the bubble divergences.
Of course this argumentation is not yet conclusive since it contain hypothesis and unresolved issues.}
From the GFT point of view the bubble divergences is analogous to the loop divergences in usual field theory.
We know that such divergences are the manifestation of a non trivial renormalization group
acting on the parameter space of field theory.

Since, as we have seen in this note,  any relevant property of spin foam model admits its dual
formulation in the GFT, this strongly suggest that a proper understanding of the action of the diffeomorphism
group on spin foam models is  related to a proper understanding of the GFT renormalization group
 that needs to be further developed \cite{Fot}.

\section{conclusions}
In this review we have seen that the GFT is a universal structure hidden behind the attempt
of dealing with quantum gravity in a background independent manner.
This attempt leads naturally to loop quantum gravity whose dynamics is governed by local spin foam models.
Such spin foam models are all expressed as  Feynman graph evaluation of a GFT.
In analogy with usual field theory whose Feynman graphs describe the dynamics of interacting relativistic
particles, this leads to the point of view that GFT is a third quantization of gravity.
If one takes this property seriously this forces us to a change of point of view in which GFT is more fundamental
than spin foam models. This has various consequences, first the classical equation of motion of group field
theory is related to the Wheeler-DeWitt equation expressed in loop variables and a natural proposal
for a triangulation independent  physical scalar product is obtained.
Second, the quantization of the GFT leads to a proposal for quantum gravity amplitudes including
sum over topologies.

We will like this letter to be an invitation for the reader to look more closely and further develop the
GFT's as a third quantized version of gravity.
As we have argued, in order to insure that these theories  effectively encode
the dynamics of General relativity one needs to gain an understanding on
the action of diffeomorphisms on spin foams model and
its counterpart in GFT, presumably implemented as a renormalisation group.
We have discussed so far pure gravity models and a consistent inclusion of matter fields and particles
in the GFT framework is clearly needed. Finally, an understanding of the physical meaning and properties
 of GFT instantons will provides us a window into the non perturbative physics of these theories.

{\bf Acknowledgments:} we would like to thank E. R. Livine, A. Starodubtsev and especially A. Perez for their useful remarks
and corrections on the manuscript.

\end{document}